# Explicit drain current, charge and capacitance model of graphene field-effect transistors

David Jiménez

Departament d'Enginyeria Electrònica, Escola d'Enginyeria,
Universitat Autònoma de Barcelona, 08193-Bellaterra, Spain

david.jimenez@uab.es

*Dated 13-May-2011*

**Abstract**

I present a compact physics-based model of the drain current, charge and capacitance of graphene field-effect transistors, of relevance for exploration of DC, AC and transient behavior of graphene based circuits. The physical framework is a field-effect model and drift-diffusion carrier transport incorporating saturation velocity effects. First, an explicit model has been derived for the drain current. Using it as a basis, explicit closed-form expressions for the charge and capacitances based on the Ward-Dutton partition scheme, covering continuosly all operation regions. The model is of special interest for analog and radio-frequency applications where bandgap engineering of graphene could be not needed.

***Index Terms-*** *graphene, field-effect transistor, analog, radio-frequency,* modeling

**Introduction**

Graphene has emerged as a material of special interest to make nanoelectronic integrated circuits beyond silicon based technology. This is due to remarkable electronic properties, like high mobility (~$10^5$ cm$^2$/V-s) and saturation velocity (~$10^8$ cm/s), together with a promising ability to scale to short gate lengths and high speeds by virtue of its thinness. Besides, advances on synthesis of large-scale graphene sheets of high quality and low cost using chemical vapor deposition techniques (CVD), are creating an appropiate framework for a new technology based on graphene to be introduced [1]. The main concern with graphene is the absence of a gap, which results in a poor ON-OFF current ratio, therefore limiting the usefulness of graphene in digital applications. However, zero gap graphene could still be very useful in analog and radiofrequency (RF) applications where high ON/OFF current ratios are not required [2]. In small signal amplifiers, for instance, the transistor is operated in the ON-state and small RF signals that are to be amplified are superimposed onto the DC gate-source voltage. Instead, what is needed to push the limits of many analog/RF figures of merit, for instance the cut-off frequency or the intrinsic gain, is an operation region where high transconductance together with a small output conductance is accomplished. These conditions are realized for state-of-the-art graphene field-effect transistors (GFETs). Specifically, for large-area GFETs, the output characteristic shows saturation like behavior that could be exploited for analog/RF applications [3]. Recently, important progress for obtaining a saturated output characteristic has been made based on CVD graphene [4]. Using GFET technology, cut-off frequencies in the THz range are envisioned [5]. It is worth noting that cut-off frequencies in the range of hundreds of GHz have been demonstrated with a non-optimized technology [1,6,7].

To boost the development of GFET technology, modeling of the electrical characteristics is essential to cover aspects as device design optimization, projection of performances, and exploration of analog / RF circuits providing new or improved functionalities [8,9].

Recently, an explicit compact model for the current-voltage (I-V) characteristics of GFET was proposed [9]. Taking this work as a basis, which provides the DC behavior, we further develop a compact physics-based model of the charge and capacitances of GFETs. This is necessary to address AC and transient simulations of graphene based circuits. The physical framework is a field-effect model and drift-diffusion carrier transport with saturation velocity effects, which is accurate in explaining the I-V behavior of GFETs [3,10]. A Ward-Dutton's charge partition scheme provides the technique to model the terminal charges together with self-capacitances and transcapacitances of GFET [11]. Explicit closed-form expressions have been derived for these quantities continuosly covering all operation regions. The model is intended to be the kernel to build up CAD simulators of graphene based circuits.

**Drain current model**

In this section we present the drain current model, which provides the basis for both the charge and capacitance models. More details can be found in Ref. [9]. Let us consider a dual gate GFET with the cross-section depicted in Fig. 1. It consists of a graphene sheet playing the role of the active channel. The source and drain electrodes are contacting the graphene channel and are assumed to be ohmic. The electrostatic modulation of the carrier concentration in graphene is achieved via a double-gate stack consisting of top and bottom gate dielectric and the corresponding metal gate. The source is grounded and considered the reference potential in the device.

The electrostatics of this device can be understood using the equivalent capacitive circuit depicted in Fig. 2 (see Ref. [10]). Here, $C_t$ and $C_b$ are the top and bottom oxide capacitances and $C_q$ represents the quantum capacitance of the graphene. The potential $V_c$ represents the voltage drop across $C_q$, and the relation between them is given by $C_q = k|V_c|$, where $k = \frac{2q^2}{\pi}\frac{q}{(\hbar v_F)^2}$ and $v_F$ (=$10^6$ m/s) is the Fermi velocity [12]. This expression is valid under the condition $qV_c \gg k_B T$. The potential $V(x)$ is the voltage drop in the graphene channel, which is zero at the source end (x=0) and equal to the drain-source voltage ($V_{ds}$) at the drain end (x=L). Applying basic circuit laws to the equivalent capacitive network and noting that the overall net mobile sheet charge density in the graphene channel, defined as $Q_c=q(p-n)$, is equal to $-(1/2)C_q V_c$, the following relation can be obtained:

$$V_c(x) = \left(V_{gs} - V_{gs0} - V(x)\right)\frac{C_t}{C_t + C_b + \frac{1}{2}C_q} + \left(V_{bs} - V_{bs0} - V(x)\right)\frac{C_b}{C_t + C_b + \frac{1}{2}C_q} \quad (1)$$

where $V_{gs}$-$V_{gs0}$ and $V_{bs}$-$V_{bs0}$ are the top and back gate-source voltage overdrive, respectively. These quantities comprise work-function differences between the gates and the graphene channel, eventual charged interface states at the graphene/oxide interfaces, and intentional or unintentional doping of the graphene.

To model the drain current of the GFET a drift-diffusion transport is assumed under the form $I_{ds}=-W|Q_c(x)|v(x)$, where W is the gate width, $|Q_c|$ is the free carrier sheet density in the channel at position x, and $v(x)$ the carrier drift velocity. The absolute value of $Q_c$ is taken because negative charged electrons and positive charged holes move in opposite directions under the electric field action, additively contributing to the drain-current. Using a soft-saturation model, consistent with Monte Carlo simulations [13], v(x) can be expressed as v=μE/(1+μ|E|/v$_{sat}$), where E is the electric field, μ is the carrier low-field mobility, and v$_{sat}$ is the saturation velocity. We assume that the saturation velocity is pinned to the Fermi velocity (v$_{sat}$=v$_F$), greatly simplifying the algebra. A more refined model, relating the saturation velocity with the carrier density, could be considered instead [9]. Applying E=-dV(x)/dx, combining the above expressions for v and v$_{sat}$, and integrating the resulting equation over the device length, the drain current becomes:

$$I_{ds} = \frac{\mu W \int_0^{V_{ds}} |Q_c| dV}{L + \mu \frac{|V_{ds}|}{v_F}} \quad (2)$$

The denominator represents an effective length (L$_{eff}$) that takes into account the saturation velocity effect. In order to get an explicit expression for the drain current, the integral in Eq. (2) is solved using V$_c$ as the integration variable and consistently expressing Q$_c$ as a function of V$_c$ :

$$I_{ds} = \frac{\mu W \int_{V_{cs}}^{V_{cd}} |Q_c(V_c)| \frac{dV}{dV_c} dV_c}{L + \mu \frac{|V_{ds}|}{v_F}} \quad (3)$$

Here V$_c$ is obtained from Eq. (1) and can be written as:

$$V_c = \frac{-(C_t + C_b) + \sqrt{(C_t + C_b)^2 \pm 2k[(V_{gs} - V_{gs0} - V)C_t + (V_{bs} - V_{bs0} - V)C_b]}}{\pm k} \quad (4)$$

where the positive (negative) sign applies when $(V_{gs} - V_{gs0} - V)C_t + (V_{bs} - V_{bs0} - V)C_b > 0 \ (< 0)$. The channel potential at the source ($V_{cs}$) is determined as $V_c(V=0)$. Similarly, the channel potential at the drain ($V_{cd}$) is determined as $V_c(V=V_{ds})$. Moreover, Eq. (1) provides the relation $\frac{dV}{dV_c} = -\left(1 + \frac{kV_c sgn(V_c)}{C_t + C_b}\right)$ entering in Eq. (3), where *sgn* refers to the sign function. On the other hand, the charge sheet density can be written as $|Q_c(V_c)| = kV_c^2/2$. Inserting these expressions into Eq. (3), the following explicit drain current expression can be finally obtained:

$$I_{ds} = \frac{\mu k}{2} \frac{W}{L_{eff}} \{g(V_c)\}_{V_{cs}}^{V_{cd}}$$

$$g(V_c) = \frac{-V_c^3}{3} - sgn(V_c)\frac{kV_c^4}{4(C_t + C_b)}$$

$$L_{eff} = L + \mu \frac{|V_{ds}|}{v_F} \quad (5)$$

Useful expressions for both the transconductance ($g_m = \frac{\partial I_{ds}}{\partial V_{gs}}$) and output conductance ($g_{ds} = \frac{\partial I_{ds}}{\partial V_{ds}}$) are given in the Appendix.

In order to test the proposed model we have benchmarked the resulting I-V characteristics with experimental results extracted from device in Ref. [14]. This is a top-gate device with L=10 μm, W=5 μm, and hafnium oxide as a dielectric with thickness of 40 nm. The flat-band voltage was $V_{gs0}$=0.85 V according to the experiment. A low-field mobility of 7500 cm$^2$/V-s for both electrons and holes and source/drain resistance of 300 Ω provides a good fit with the experiment.

To reproduce any experimental current-voltage characteristics of GFETs, accounting of the voltage drop at the source/drain contacts is necessary. This quantity must be removed from the external applied bias ($V_{ds\_ext}$) in order to get the internal $V_{ds}$. This is simply done by solving the equation $V_{ds} = V_{ds\_ext} - I_{ds}(V_{ds})(R_s + R_d)$, where $R_s$ and $R_d$ represents the source and drain contact resistance, respectively. Finally note that $I_{ds}$ is a function of $V_{ds}$ as given by Eq. (5).

The resulting I-V characteristics are shown in Fig. 3. We have extended the analyzed drain-source voltage range beyond the experiment to show the predictive behavior of the model. The transfer characteristics exhibit the expected ambipolar behavior dominated by holes (electrons) below and above the Dirac gate voltage. The output characteristics show a saturation-like behavior with a second linear region. The agreement between the proposed model (solid lines) and the experiment (symbols) demonstrates its accuracy. A closer agreement can even be obtained using a more refined model given in Ref. [9] (dashed lines).

**Charge model**

In this section we provide analytical and continuous expressions of charges associated with each terminal. We consider the important case of a double-gate configuration, where the two gates are connected together. This allows considering the GFET as a 3-terminal device, therefore making the analysis simpler. Extension to 4-terminal devices could be done using the same principles outlined in this work.

The terminal charges $Q_g$, $Q_d$, and $Q_s$, associated with the gate, drain and source electrodes have to be considered. The gate charge can be computed by integrating the net mobile carrier concentration along the channel. On the other hand, the drain and source charges can be computed based on a Ward-Dutton's linear charge partition scheme, which preserves the charge conservation [11].

$$Q_g = -W \int_0^L Q_c(y) dy$$

$$Q_d = W \int_0^L \frac{y}{L} Q_c(y) dy$$

$$Q_s = -(Q_g + Q_d) \quad (6)$$

These expressions can be conveniently written using $V_c$ as the integration variable, as it was done to model the drain-current. There, the channel charge density is given by $Q_c(V_c) = -sgn(V_c) k V_c^2 / 2$. A negative (positive) $V_c$ means that the channel charge density is dominated by holes (electrons). Notice that the absolute value of the channel charge density, needed for the drain-current calculation, is absent here. As a consequence, integration of this quantity over the channel length could result in cancelation of the terminal charges. This situation physically would correspond to the appearance of pinch-off at some point in the channel. On the other hand, according to current continuity equation, $dy = -\mu W |Q_c(y)| \frac{dV}{I_{ds}(y)}$ and y is equal to $\frac{L_{eff}(g(V_{cs}) - g(V_c))}{g(V_{cd}) - g(V_{cs})}$ based on the fact that the drain-current is the same at any point *y* in the channel. Substituting $Q_c$, dy, and y into Eq. (6) allow integrals to be expressed as

$$Q_g = -W \int_{V_{cs}}^{V_{cd}} Q_c(V_c) \left( \frac{-\mu W |Q_c(V_c)|}{I_{ds}} \right) \frac{dV}{dV_c} dV_c$$

$$Q_d = W \int_{V_{cs}}^{V_{cd}} \frac{1}{L} \left( \frac{L_{eff}(g(V_{cs}) - g(V_c))}{g(V_{cd}) - g(V_{cs})} \right) \left( \frac{-\mu W |Q_c(V_c)|}{I_{ds}} \right) Q_c(V_c) \frac{dV}{dV_c} dV_c \quad (7)$$

These integrals have been carried out to yield the charge model

$$Q_g = k \frac{WL_{eff}}{2} \frac{\{f(V_c)\}_{V_{cs}}^{V_{cd}}}{\{g(V_c)\}_{V_{cs}}^{V_{cd}}}$$

$$Q_d = -k \frac{WL_{eff}^2}{2L} \frac{g(V_{cs})\{f(V_c)\}_{V_{cs}}^{V_{cd}} + \{m(V_c)\}_{V_{cs}}^{V_{cd}}}{\left[\{g(V_c)\}_{V_{cs}}^{V_{cd}}\right]^2}$$

$$Q_s = -(Q_g + Q_d) \quad (8)$$

where

$$f(V_c) = sgn(V_c)\frac{V_c^5}{5} + \frac{kV_c^6}{6(C_t + C_b)}$$

$$m(V_c) = sgn(V_c)\frac{V_c^8}{24} + \frac{7kV_c^9}{108(C_t + C_b)} + sgn(V_c)\frac{k^2 V_c^{10}}{40(C_t + C_b)^2} \quad (9)$$

These expressions apply to a long-channel double-gate GFET incorporating saturation velocity effects as given by (5).

**Capacitance model**

In this section we present explicit expressions for the self-capacitances and transcapacitances of the GFET. For the 3-terminal case under consideration it results in nine nonreciprocal capacitances for transient or small-signal simulation. These are defined as

$$C_{vw} = -\frac{\partial Q_v}{\partial V_w} \quad v \neq w$$

$$C_{vv} = \frac{\partial Q_v}{\partial V_v} \quad otherwise \quad (10)$$

where *v* and *w* stand for *g*, *s*, and *d*. Applying the charge conservation law, the following relation between capacitances can be obtained [15]:

$$C_{ss} = C_{sd} + C_{sg} = C_{ds} + C_{gs}$$
$$C_{gg} = C_{gs} + C_{gd} = C_{sg} + C_{dg}$$
$$C_{dd} = C_{ds} + C_{dg} = C_{sd} + C_{gd} \quad (11)$$

This relation leaves only four independent capacitances, namely: (1) $C_{gd}$, (2) $C_{dd}$, (3) $C_{dg}$, and (4) $C_{gg}$, which can be derived using the previous charge model given by Eqs. (8)-(9).

$$C_{gg} = \frac{\partial Q_g}{\partial V_{cd}} \times \frac{\partial V_{cd}}{\partial V_g} + \frac{\partial Q_g}{\partial V_{cs}} \times \frac{\partial V_{cs}}{\partial V_g}$$

$$C_{gd} = -\frac{\partial Q_g}{\partial V_{cd}} \times \frac{\partial V_{cd}}{\partial V_d}$$

$$C_{dd} = \frac{\partial Q_d}{\partial V_{cd}} \times \frac{\partial V_{cd}}{\partial V_d}$$

$$C_{dg} = -\frac{\partial Q_d}{\partial V_{cd}} \times \frac{\partial V_{cd}}{\partial V_g} - \frac{\partial Q_d}{\partial V_{cs}} \times \frac{\partial V_{cs}}{\partial V_g} \quad (12)$$

where

$$\frac{\partial Q_g}{\partial V_{cd}} = k \frac{WL_{eff}}{2} \left\{ \frac{-g'(V_{cd})\{f(V_c)\}_{V_{cs}}^{V_{cd}}}{\left[\{g(V_c)\}_{V_{cs}}^{V_{cd}}\right]^2} + \frac{h(V_{cd})}{\{g(V_c)\}_{V_{cs}}^{V_{cd}}} \right\}$$

$$\frac{\partial Q_g}{\partial V_{cs}} = k \frac{WL_{eff}}{2} \left\{ \frac{g'(V_{cs})\{f(V_c)\}_{V_{cs}}^{V_{cd}}}{\left[\{g(V_c)\}_{V_{cs}}^{V_{cd}}\right]^2} - \frac{h(V_{cs})}{\{g(V_c)\}_{V_{cs}}^{V_{cd}}} \right\}$$

$$\frac{\partial Q_d}{\partial V_{cd}} = -k\frac{WL_{eff}^2}{2L}\left\{\frac{-2g'(V_{cd})}{\left[\{g(V_c)\}_{V_{cs}}^{V_{cd}}\right]^3}\left(g(V_{cs})\{f(V_c)\}_{V_{cs}}^{V_{cd}} + \{m(V_c)\}_{V_{cs}}^{V_{cd}}\right)\right.$$

$$\left. + \frac{1}{\left[\{g(V_c)\}_{V_{cs}}^{V_{cd}}\right]^2}\left(g(V_{cs})h(V_{cd}) + n(V_{cd})\right)\right\}$$

$$\frac{\partial Q_d}{\partial V_{cs}} = -k\frac{WL_{eff}^2}{2L}\left\{\frac{2g'(V_{cs})}{\left[\{g(V_c)\}_{V_{cs}}^{V_{cd}}\right]^3}\left(g(V_{cs})\{f(V_c)\}_{V_{cs}}^{V_{cd}} + \{m(V_c)\}_{V_{cs}}^{V_{cd}}\right)\right.$$

$$\left. + \frac{1}{\left[\{g(V_c)\}_{V_{cs}}^{V_{cd}}\right]^2}\left(g'(V_{cs})\{f(V_c)\}_{V_{cs}}^{V_{cd}} - g(V_{cs})h(V_{cs}) - n(V_{cs})\right)\right\}$$

$$\frac{\partial V_{cd}}{\partial V_g} = -\frac{\partial V_{cd}}{\partial V_d} = \left(1 + sgn(V_{cd})\frac{kV_{cd}}{C_t + C_b}\right)^{-1}$$

$$\frac{\partial V_{cs}}{\partial V_g} = \left(1 + sgn(V_{cs})\frac{kV_{cs}}{C_t + C_b}\right)^{-1}$$

$$g'(V_c) = -V_c^2 - sgn(V_c)\frac{kV_c^3}{C_t + C_b}$$

$$h(V_c) = sgn(V_c)V_c^4 + \frac{kV_c^5}{C_t + C_b}$$

$$n(V_c) = sgn(V_c)\frac{V_c^7}{3} + \frac{7kV_c^8}{12(C_t + C_b)} + sgn(V_c)\frac{k^2 V_c^9}{4(C_t + C_b)^2} \quad (13)$$

In the derivation of the capacitances, it should be noted that $\frac{\partial V_{cd}}{\partial V_s} = \frac{\partial V_{cs}}{\partial V_d} = 0$ because the voltage applied to source has no control over the charge at drain, and vice versa [16].

**Model assessment with discussion of the operation regions**

To illustrate the model outcome, Fig. 4 shows the terminal charges and capacitances for the examined device (Ref. [14]) as a function of $V_{gs}$, where $V_{ds}$ has been fixed (=-0.5V). An insight of the terminal charge evolution can be obtained by looking at the quantities $V_{cs}$ and $V_{cd}$, informing of the type of carriers that dominate at the source and drain end, respectively. The examined case corresponds to $V_{cd}>V_{cs}$ whatever is the $V_{gs}$ value. With regard to the n-type or p-type character of the channel three different situations can be distinguished depending on the gate voltage, namely:

(1) <u>Hole channel</u> ($V_{gs}<V_{gs,p\text{-off-}d}$), where $V_{gs,p\text{-off-}d}=V_{ds}+V_{gs0}$ (=0.35 V) labels the gate voltage that exactly places the pinch-off point at x=L (Fig. 4b). In this region the channel is entirely p-type because both $V_{cs},V_{cd} < 0$. As a consequence, both $Q_g,Q_d > 0$, as can be observed in Fig. 4a.

(2) <u>Ambipolar channel</u> ($V_{gs,p\text{-off-}d}<V_{gs}<V_{gs,p\text{-off-}s}$), where $V_{gs,p\text{-off-}s}=V_{gs0}$ (=0.85 V) labels the gate voltage that places the pinch-off point at x=0. In this region the channel has a mixed character, being p-type at the source end and n-type close to the drain end, respectively, consistently with $V_{cs}<0$ and $V_{cd}>0$. At some specific $V_{gs,D}=V_{gs0}+V_{ds}/2$ (=0.65 V), belonging to this interval, the condition $V_{cs}=-V_{cd}$ is satisfied shifting the pinch-off point to x=L/2. For this particular gate voltage, named Dirac voltage, electron and hole charges are exactly balanced over the channel length, resulting in $Q_g=0$. Note, however, that $Q_d$ goes across zero at $V_{gs}$ slightly smaller than $V_{gs,D}$. The reason being that $Q_d$, defined from a Ward-Dutton's linear charge partition scheme, weights the carriers at the drain-side (electrons) more than carriers at the source-side (holes).

(3) <u>Electron channel</u> ($V_{gs}>V_{gs,p\text{-off-s}}$). In this region the channel is entirely n-type because both $V_{cs}, V_{cd} > 0$, consistently with the quantities $Q_g, Q_d < 0$.

It is worth noting that if the selected $V_{ds}$ results in $V_{cs} > V_{cd}$ the operation regions would be instead: (1) <u>Hole channel</u> ($V_{gs}<V_{gs,p\text{-off-s}}$), (2) <u>Ambipolar channel</u> ($V_{gs,p\text{-off-s}}<V_{gs}<V_{gs,p\text{-off-d}}$), (3) <u>Electron channel</u> ($V_{gs}>V_{gs,p\text{-off-d}}$).

Next I examine the capacitances of the GFET under test (Fig. 4c). Interestingly, the self-capacitance -$C_{gg}$ is symmetric around the Dirac voltage, where a maximum value arises. It also presents two local minimum at $V_{gs,p\text{-off-d}}$ and $V_{gs,p\text{-off-s}}$, respectively. The transcapacitances -$C_{sg}$ and $C_{dg}$ show a similar behavior, although the curves are slightly asymmetric due to the position-dependent weighting of the mobile charge in the Ward-Dutton's partition scheme. Moreover, the position of the maximum is slightly shifted respect to the Dirac voltage. On the other hand, the capacitances $C_{sd}$, -$C_{gd}$, $C_{dd}$ (-$C_{ss}$, -$C_{gs}$, -$C_{ds}$) exhibit a zero value at $V_{gs,p\text{-off-d}}$ ($V_{gs,p\text{-off-s}}$) and a maximum around $V_{gs,D}$.

Next, we examine the terminal charges and capacitances as a function of $V_{ds}$ for a fixed $V_{gs}$ (=-0.25 V) (see Fig. 5). In this case $V_{cs}$ is constant (<0) and different operation regions appear driven by $V_{cd}$. Two different channel types are then possible according to the drain voltage, namely:

(1) <u>Ambipolar channel</u> ($V_{ds}<V_{ds,p\text{-off-d}}$), where $V_{ds,p\text{-off-d}}=V_{gs}-V_{gs0}$ (=-1.1V) labels the drain voltage that places the pinch-off point at x=L. In this region the channel has a mixed p-type and n-type character at the source and drain end, respectively, consistently with $V_{cs}<0$ and $V_{cd}>0$. On the other hand, $Q_g<0$ as long as $V_{ds}<V_{ds,D}$, where $V_{ds,D} = 2\ V_{ds,p\text{-off-d}}$ (=-2.2 V) is the drain voltage that places the pinch-off just in the middle of the channel (x=L/2), where the relation $V_{cs}=-V_{cd}$ is fulfilled. Regarding the drain-current, this is roughly linear with $V_{ds}$. This region is named as **second linear region**. As long as $V_{ds}$ approaches $V_{ds,p\text{-off-d}}$ the output characteristics get **saturated**. This behavior can be observed in the GFET output characteristics of the device under test in Fig. 3b. Note that $V_{ds\_ext}$ has been considered as the independent variable for plotting this figure, being different from $V_{ds}$; the later meaning the internal drain-source voltage.

(2) Hole channel ($V_{ds}>V_{ds,p\text{-off-d}}$). In this region the channel is entirely p-type, consistently with $V_{cs},V_{cd}<0$ and $Q_g,Q_d>0$. For this particular range of $V_{ds}$, the output characteristic is linear with $V_{ds}$ and it is named as **first linear region**.

It is worth noting that if the selected $V_{gs}$ results in $V_{cs}>0$ the operation regions would be instead: (1) Electron channel ($V_{ds}<V_{ds,p\text{-off-d}}$), (2) Ambipolar channel ($V_{ds}>V_{gs,p\text{-off-d}}$).

The capacitances $C_{sd}$, $-C_{gd}$, and $C_{dd}$ as a function of $V_{ds}$ show a maximum around $V_{ds,D}$ and a minimum zero capacitance at $V_{ds,p\text{-off-d}}$, where all the terminal charges $Q_s$, $Q_g$, and $Q_d$ get saturated (Figs. 5b and 5c).

**Conclusion**

In conclusion, I have presented an explicit and compact drain current, charge and capacitance model for GFETs based on a field-effect model and drift-diffusion carrier transport, including saturation velocity effects. The model captures the physics of all operation regions within a single expression for the drain current and each terminal charge and capacitance. It is of especial interest as a tool for design of analog and RF applications [17]. Additional physical effects as, for example, short-channel effects, non-quasi static effects, extrinsic capacitances [5], and mobility model [3,18] need to be incorporated into the long-channel core presented here to build a complete GFET compact model.

## Acknowledgments

We acknowledge the funding of the Ministerio de Ciencia e Innovación under contracts FR2009-0020 and TEC2009-09350, and the DURSI of the Generalitat de Catalunya under contract 2009SGR783.

## Appendix

In this section useful expressions for both $g_m$ and $g_{ds}$ are provided.

$$g_m = \frac{\partial I_{ds}}{\partial V_{gs}} = \frac{\partial I_{ds}}{\partial V_{cd}} \times \frac{\partial V_{cd}}{\partial V_{gs}} + \frac{\partial I_{ds}}{\partial V_{cs}} \times \frac{\partial V_{cs}}{\partial V_{gs}}$$

$$= \frac{\mu k}{2} \frac{W}{L_{eff}} \left( \frac{g'(V_{cd})}{1 + sgn(V_{cd})\frac{kV_{cd}}{C_t + C_b}} - \frac{g'(V_{cs})}{1 + sgn(V_{cs})\frac{kV_{cs}}{C_t + C_b}} \right) \quad (A1)$$

$$g_{ds} = \frac{\partial I_{ds}}{\partial V_{ds}} = \frac{\partial I_{ds}}{\partial V_{cd}} \times \frac{\partial V_{cd}}{\partial V_{ds}} + \frac{\partial I_{ds}}{\partial V_{cs}} \times \frac{\partial V_{cs}}{\partial V_{ds}} = -\frac{\mu k}{2} \frac{W}{L_{eff}} \frac{g'(V_{cd})}{1 + sgn(V_{cd})\frac{kV_{cd}}{C_t + C_b}} \quad (A2)$$

**Figure captions:**

Fig. 1. Cross section of the dual-gate GFET.

Fig. 2. Equivalent capacitive circuit of the dual-gate GFET.

Fig. 3. Transfer (a) and output (b) characteristics obtained from the analytical model (solid lines) compared with experimental results from Ref. [14] (symbols) and a more elaborated analytical model (dashed lines), which incorporates the effect of doping and a saturation velocity model that takes into account the carrier density (see Ref. [9]) .

Fig. 4. Terminal charges (a), channel voltage drop (b), and capacitances (c) for the examined device, as a function of the gate voltage. The channel is shown to be p-type, mixed p and n-type, or p-type, depending on the gate voltage.

Fig. 5. Terminal charges (a), channel voltage drop (b), and capacitances (c) for the examined device, as a function of the drain voltage. The channel is shown to be mixed p and n type, or p-type, depending on the drain voltage.

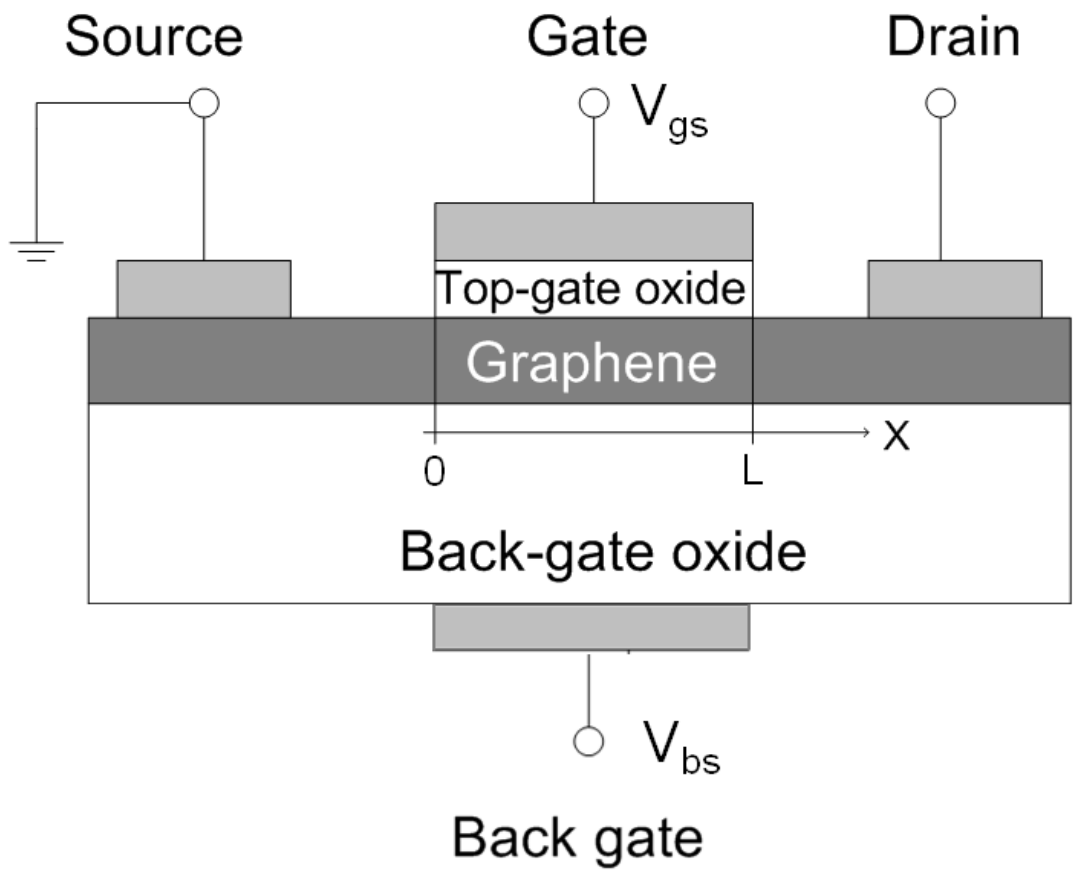

Figure 1

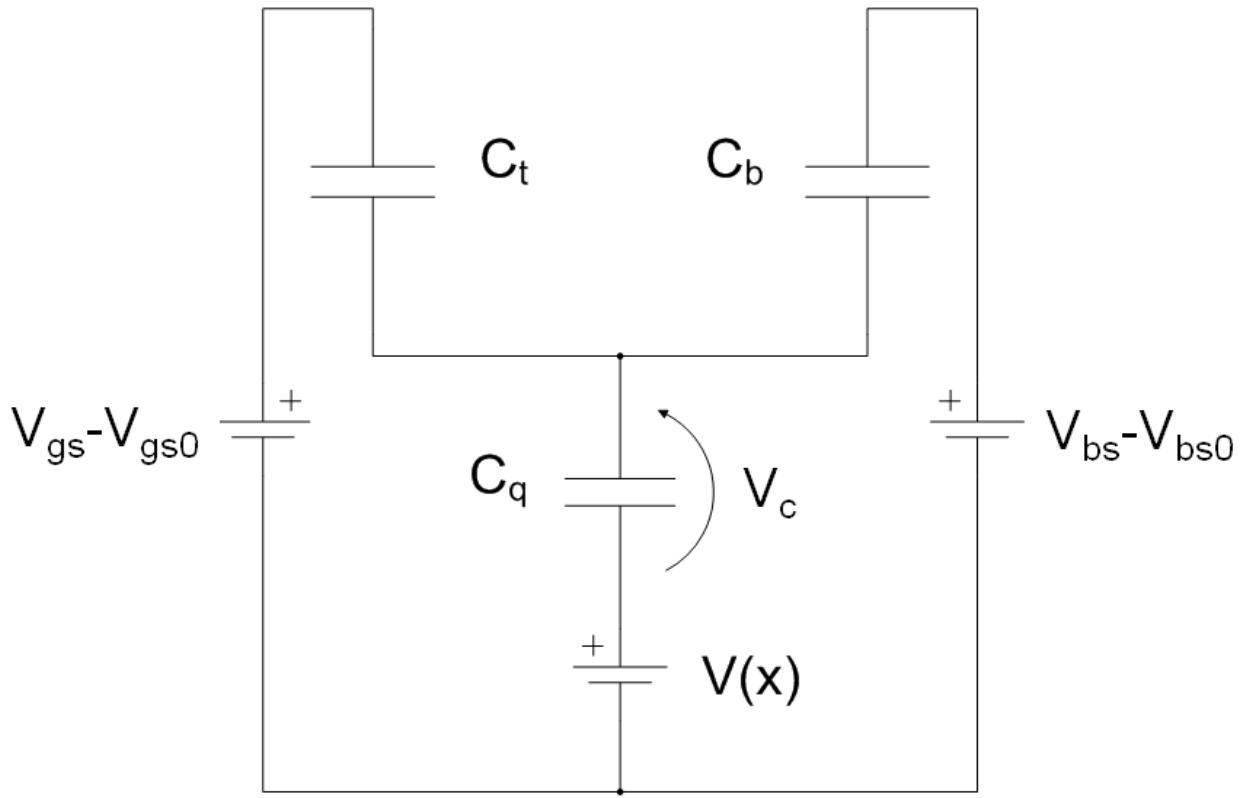

Figure 2

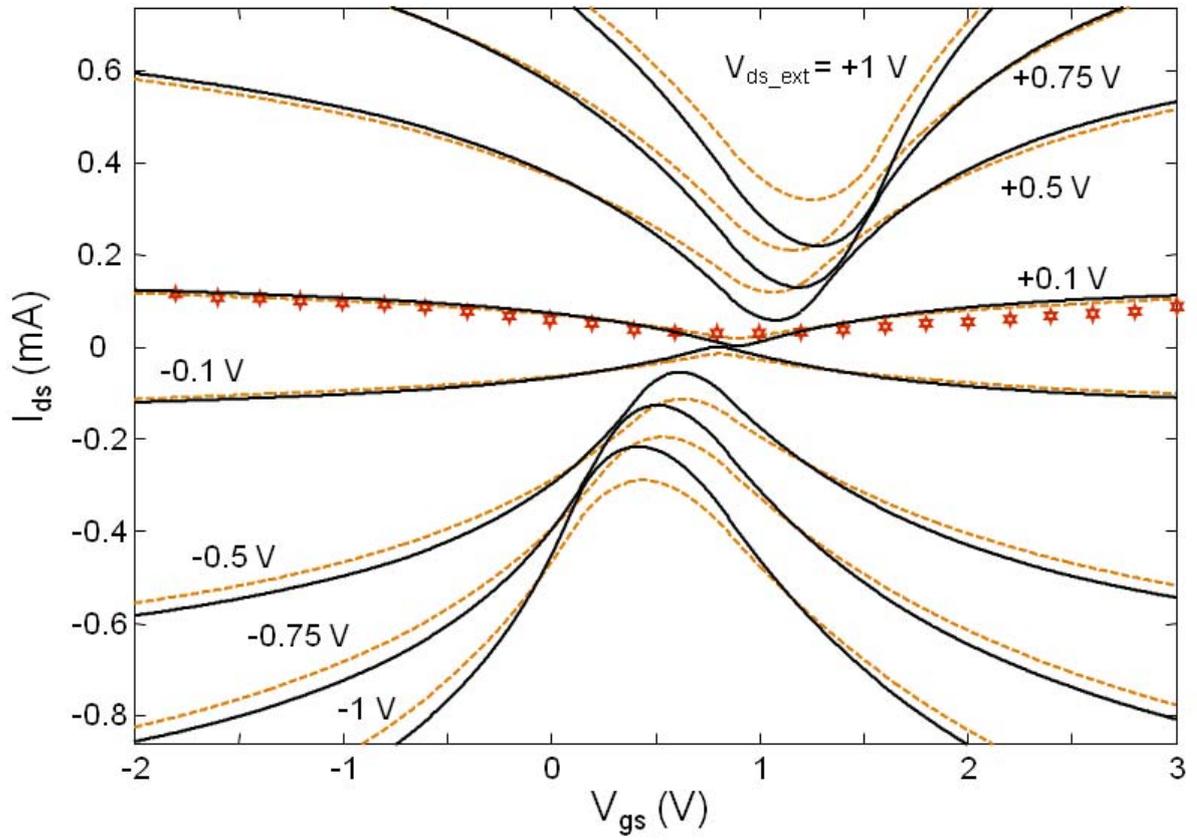

Figure 3a

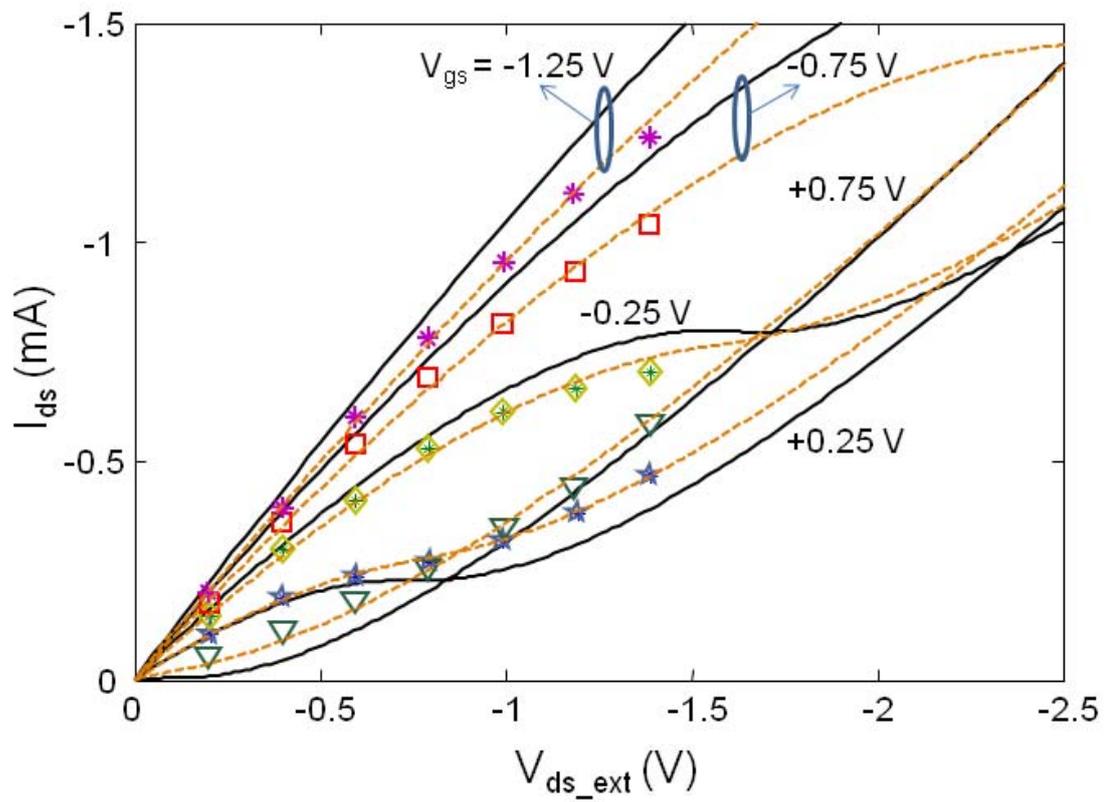

Figure 3b

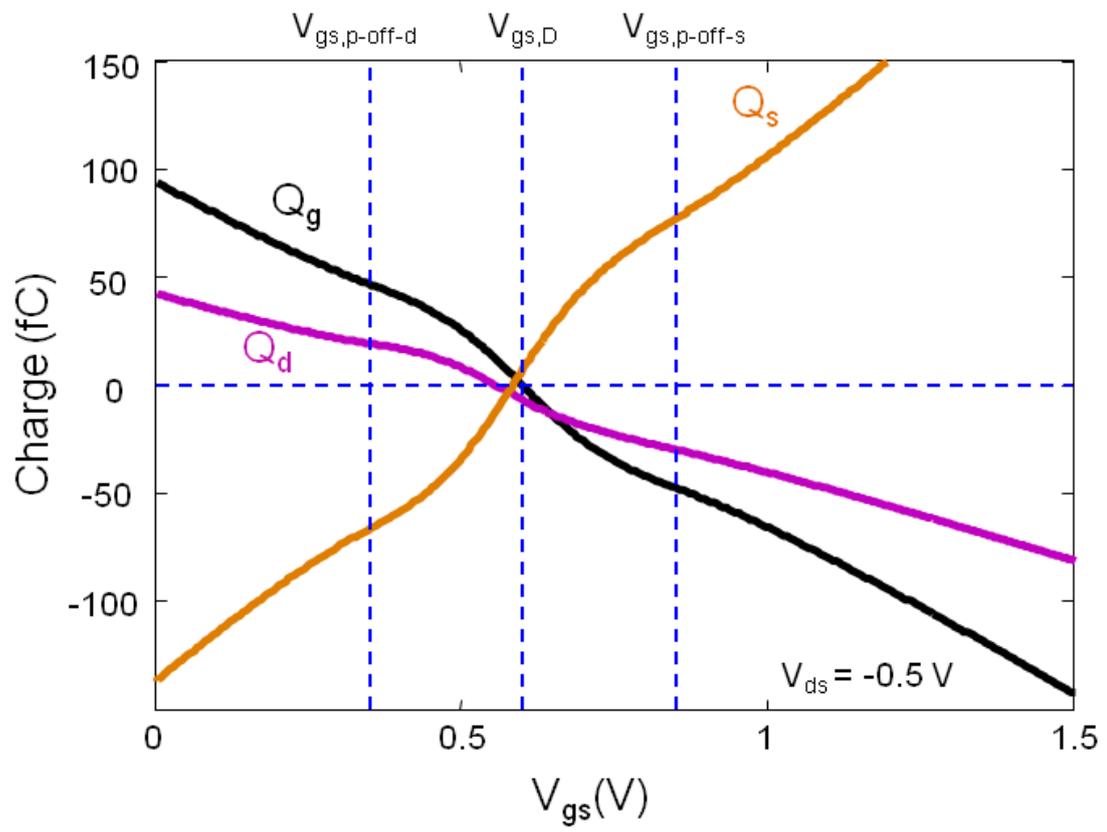

Figure 4a

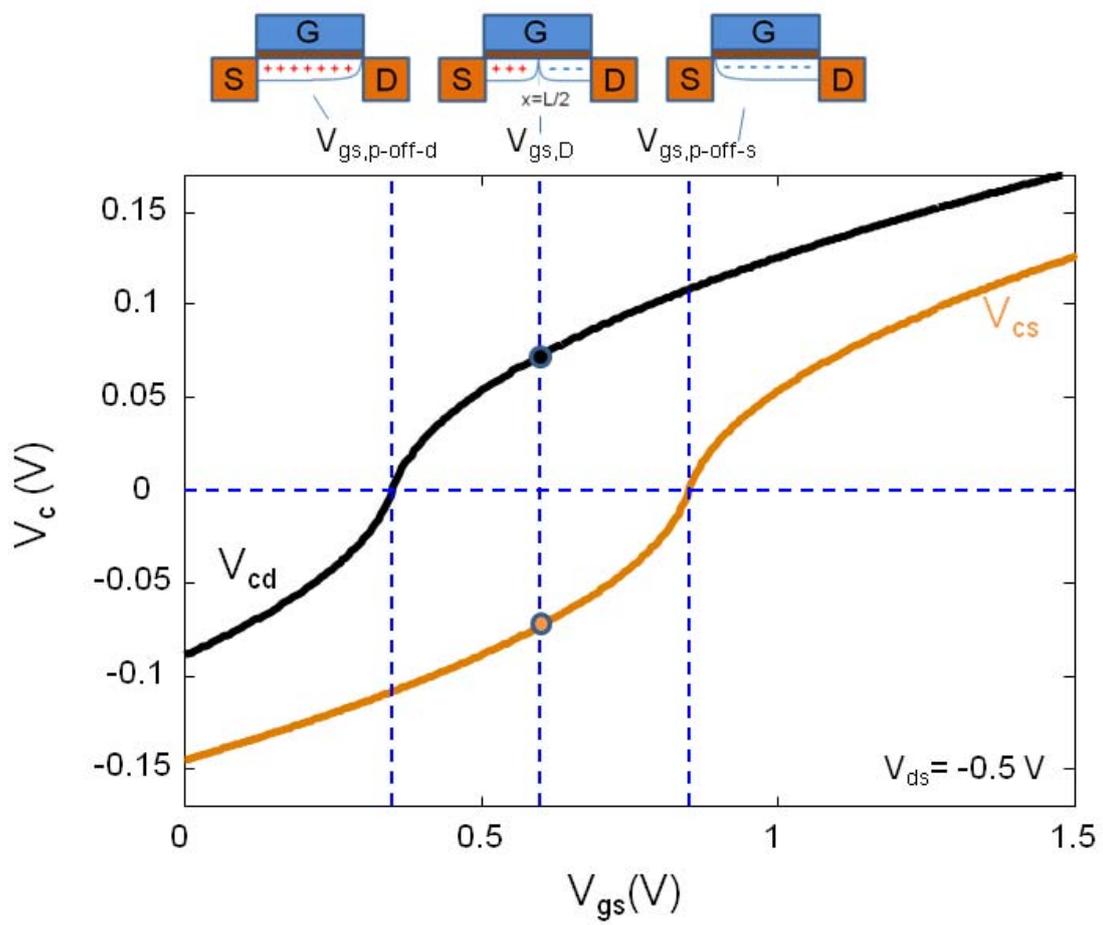

Figure 4b

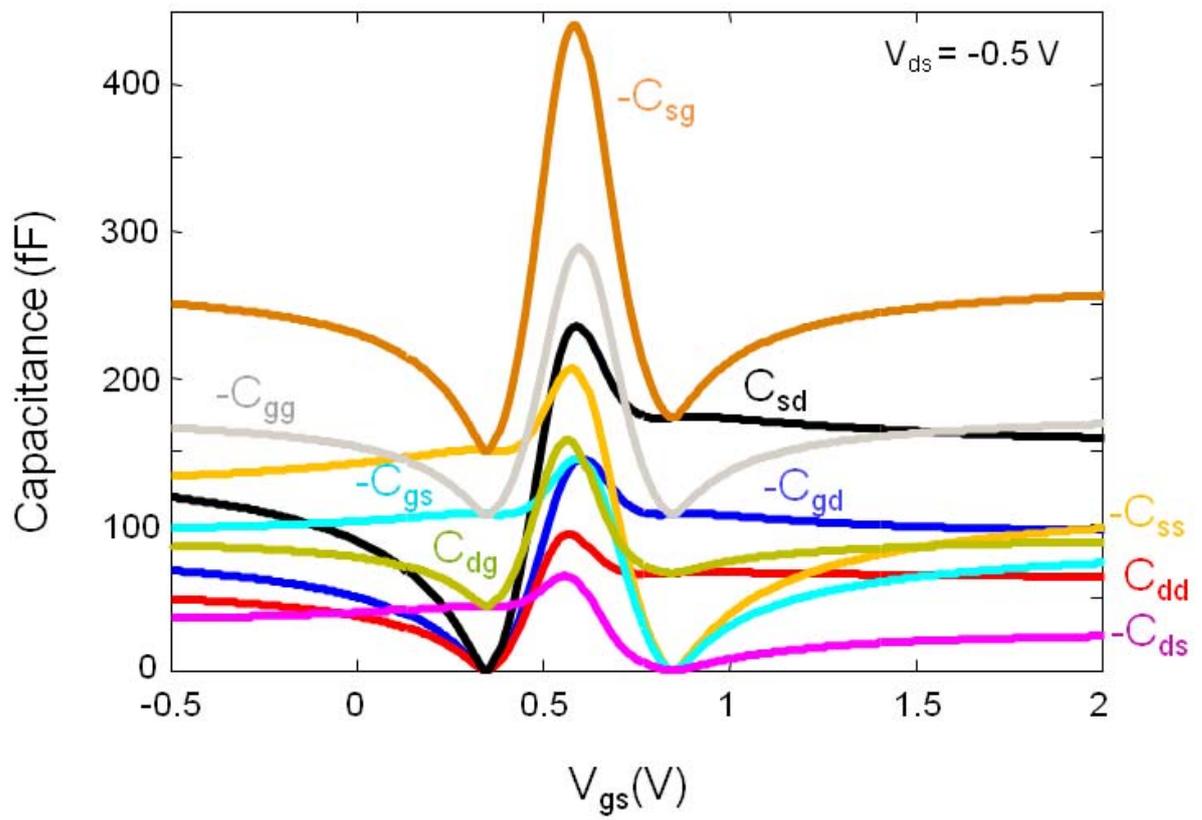

Figure 4c

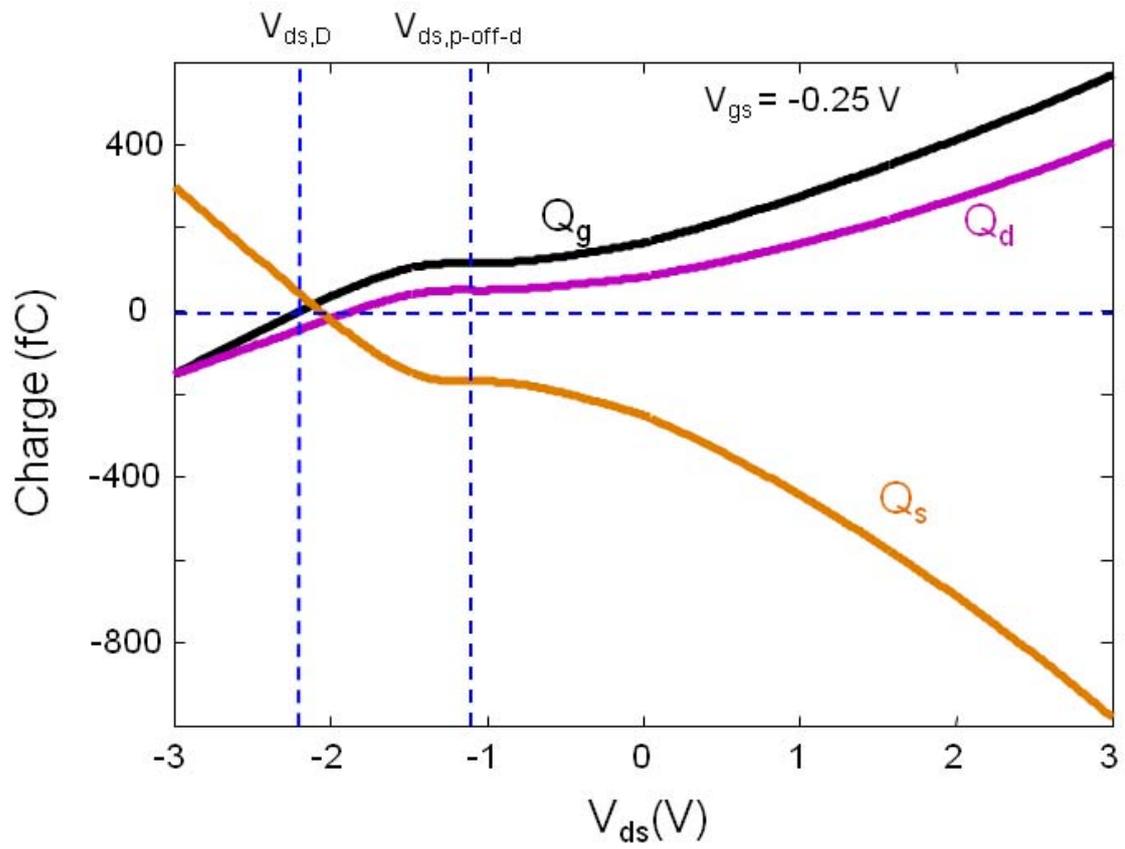

Figure 5a

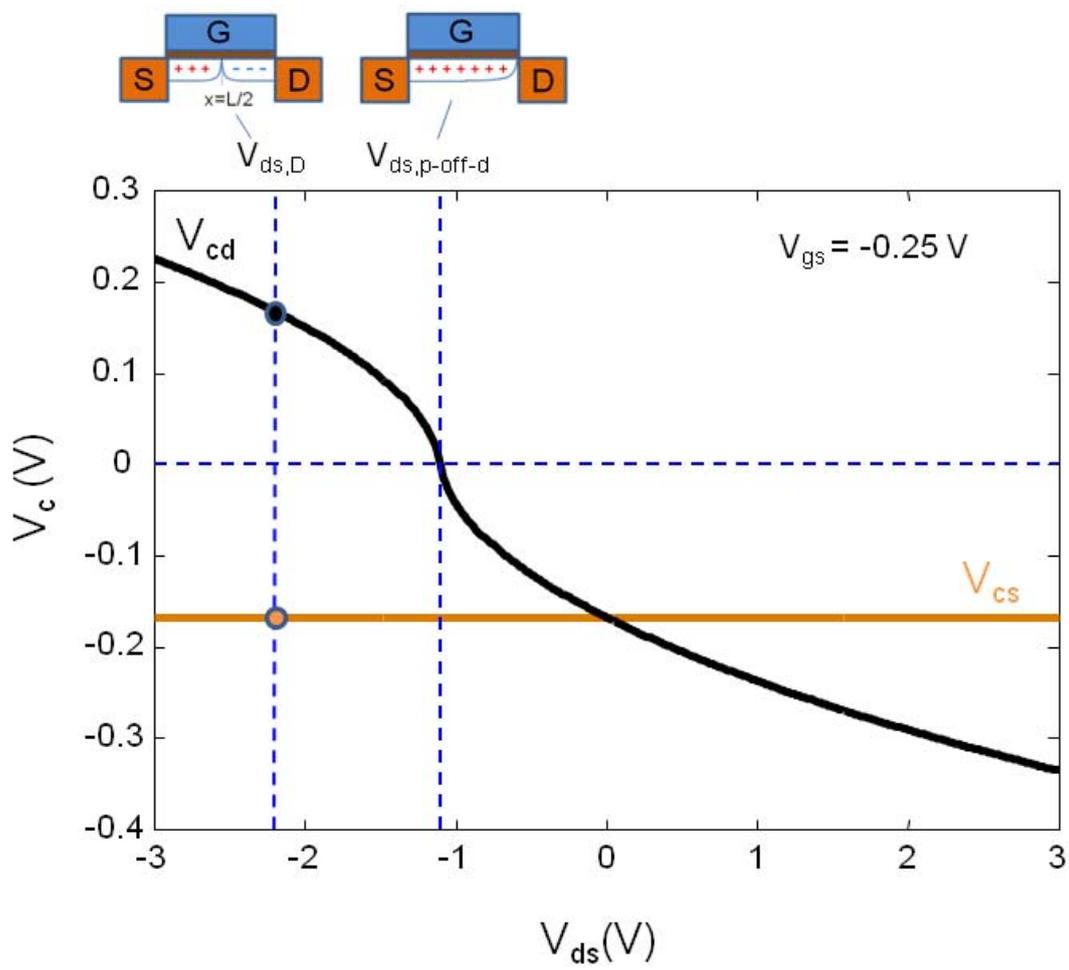

Figure 5b

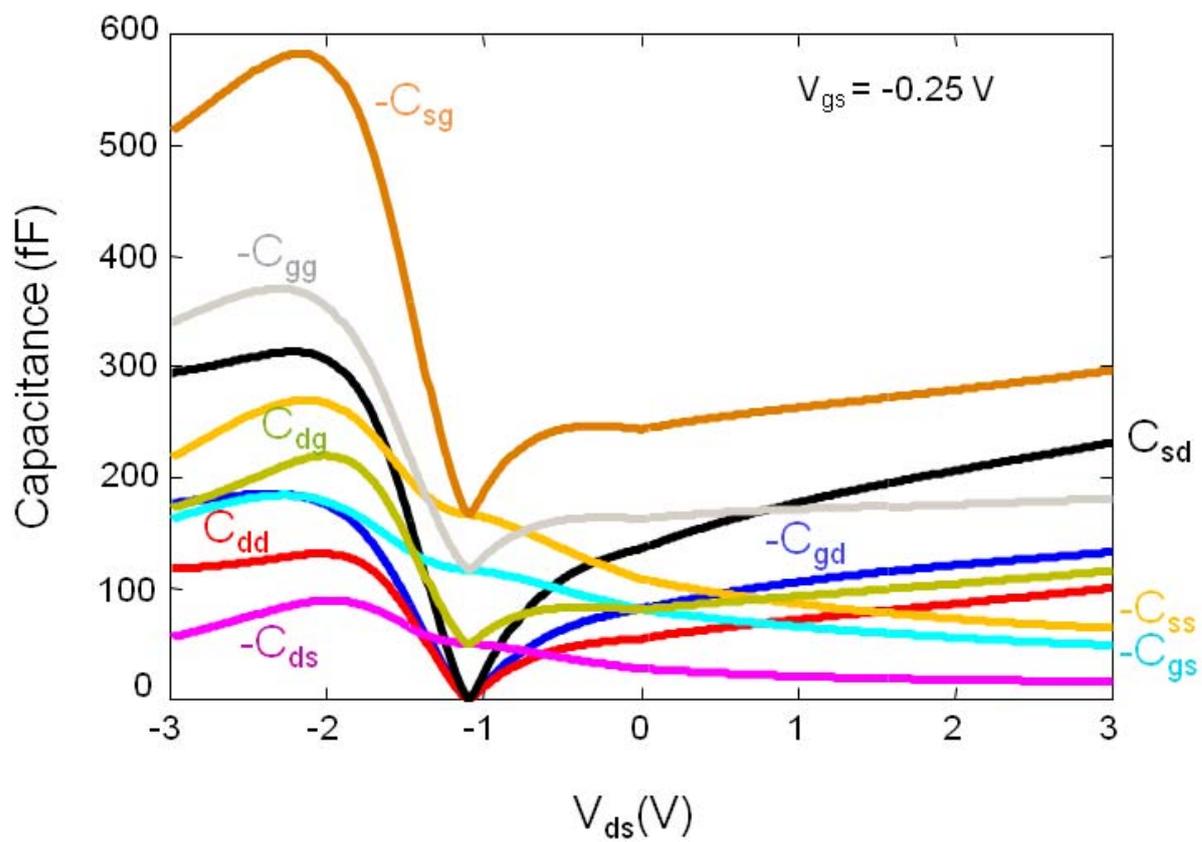

Figure 5c